\documentclass[10pt,aps,prd,twocolumn,showpacs,superscriptaddress,eqsecnum,longbibliography,nofootinbib]{revtex4-2}
\usepackage{mathrsfs,amsmath,amsthm,latexsym,amssymb,amsfonts,epsfig,cancel,enumerate,graphicx,subcaption,txfonts}
\usepackage{multirow}
\usepackage{xcolor}
\definecolor{navy}{RGB}{0,0,150}
\usepackage[colorlinks, linkcolor=blue, citecolor=blue, urlcolor=blue, plainpages=false, pdfstartview=FitH]{hyperref}
\usepackage{appendix}
\allowdisplaybreaks
\newcommand{\GZU}{School of Physics, Guizhou University, Guiyang 550025, China}

\begin{document}

\title{Periodic orbits and gravitational waveforms in quantum-corrected black hole spacetimes}

\author{Jiawei Chen}
\email{gs.chenjw23@gzu.edu.cn}
\affiliation{\GZU}

\author{Jinsong Yang}
\thanks{Corresponding author}
\email{jsyang@gzu.edu.cn}
\affiliation{\GZU}

\begin{abstract}
In this paper, we study the periodic orbits of massive particles around two quantum-corrected black holes proposed in effective quantum gravity, and explore the quantum gravity effect on both the particle orbits and the associated gravitational wave signals. First, we analyze the geodesic motion of the massive particle around the black holes. We then study two important types of bound orbits of the massive particles, the marginally bound orbit and the innermost stable circular orbit. We find that, for the first black hole, increasing the quantum parameter $\zeta$ leads to larger orbital radii and reduced angular momenta for both orbits. In contrast, the second black hole shows $\zeta$-independent orbital radii and angular momenta. By analyzing the effective potential, we determine the allowed range of the energy and the angular momentum for bound orbits, with $\zeta$-dependence only for the first black hole. We further investigate periodic orbits with a fixed energy for both black holes, revealing that the parameter $\zeta$ similarly affects the orbits, although its effect is negligible in the second black hole. Finally, we calculate the gravitational waves emitted by the periodic orbits. The results demonstrate that increasing $\zeta$ leads to a significant phase delay for the first black hole, while only inducing a subtle phase advance for the second one. Therefore, we conclude that the first black hole can be distinguished from the Schwarzschild one through gravitational wave observations, whereas the second one cannot be effectively distinguished when the quantum correction is weak.
\end{abstract}

\maketitle

\section{Introduction}

The advent of general relativity (GR) significantly deepened our understanding of gravity, with its predictions of gravitational waves (GWs) and black holes (BHs) now firmly supported by observational evidence~\cite{LIGOScientific:2016aoc,EventHorizonTelescope:2019dse,EventHorizonTelescope:2022wkp}. According to the singularity theorems~\cite{Penrose:1964wq,Hawking:1970zqf}, the gravitational collapse of massive stars in GR inevitably leads to spacetime singularities. Combined with its incompatibility with quantum mechanics, this suggests that GR may break down in extreme regimes. To address these issues, various approaches to quantum gravity have been proposed to describe the gravitational phenomena in such extreme domains~\cite{Hartle:1983ai,Strominger:1996sh,Bojowald:2008zzb,Ashtekar:2013hs}. Among them, loop quantum gravity (LQG), as a non-perturbative approach to quantum gravity, has attracted considerable attention and led to fruitful developments~\cite{Rovelli:1997yv,Thiemann:2007pyv,Ashtekar:2004eh,Ashtekar:2005qt,Han:2005km,Yang:2009fp,Perez:2012wv,Yang:2016kia,Perez:2017cmj,Ashtekar:2018lag,Zhang:2020smo,Long:2020agv,Sartini:2020ycs,Lewandowski:2022zce,Zhang:2022vsl}.

Although LQG has made remarkable progress in resolving classical singularities, whether its canonical formulation preserves covariance after the introduction of quantum parameters has remained unclear~\cite{Bojowald:2015zha}, prompting extensive efforts to address this issue~\cite{Bojowald:2015sta,BenAchour:2017jof,Bojowald:2020unm,Gambini:2022dec,Han:2022rsx}. Recently, within the Hamiltonian framework of effective quantum gravity inspired by LQG, the conditions required for covariance have been rigorously derived in static and spherically symmetric case, and some covariant quantum-corrected BH solutions characterized by a quantum parameter $\zeta$ have been carried out~\cite{Zhang:2024khj,Zhang:2024ney}. This approach has subsequently been extended to electromagnetic vacuum spacetimes with a cosmological constant~\cite{Yang:2025ufs}. These covariant quantum-corrected BH solutions have drawn great interest, leading to a number of studies investigating their physical properties~\cite{Konoplya:2024lch,Liu:2024soc,Liu:2024wal,Zhu:2024wic,Wang:2024iwt,Ban:2024qsa,Lin:2024beb,Shu:2024tut,Liu:2024iec,Bojowald:2024ium,Konoplya:2025hgp,Chen:2025ifv,Lutfuoglu:2025hwh,Al-Badawi:2025rcq,Li:2024afr,Malik:2024elk,Liu:2024wal,Malik:2024nhy,Heidari:2024bkm,Wang:2024iwt,Du:2024ujg,Shu:2024tut,Liu:2024iec,Cafaro:2024lre,Bojowald:2024ium,DelAguila:2025pgy,Konoplya:2025hgp,Xamidov:2025oqx,Lutfuoglu:2025hwh,Motaharfar:2025ihv}.

The strong gravitational field of BHs offers an important platform for testing various theories of gravity. By studying the null geodesics of massless particles (photons) near BHs, such as calculating the deflection angle of light and the BH shadow~\cite{Kudo:2024aak,Gao:2024ejs,Igata:2025plb,Kumar:2019ohr,Yang:2022btw,Wang:2023vcv}, one can extract information about BH spacetimes, and distinguish between different gravitational theories. On the other hand, the timelike geodesics of massive particles around BHs can also provide valuable insight into the BH geometry. Typically, a stellar-mass test particle orbiting a central supermassive BH forms an extreme mass ratio inspiral (EMRI) system, which serves as a potential source of GWs~\cite{Hughes:2000ssa}. The GW radiation emitted from such orbital motion carries information about both the trajectory and the background spacetime, offering a promising approach for testing different gravitational theories~\cite{Babak:2017tow}. Therefore, the study of particle orbits is of great significance for probing the spacetime geometry in strong-field regimes. Among various types of massive particle trajectories, periodic orbits represent an important class of bound orbits that play a key role in GW radiation. A massive particle on a periodic orbit around a BH returns to its initial position within a finite time. These orbits can be characterized by three integers $(z, w, v)$, which respectively describe the zoom, whirl, and vertex behavior of the trajectory~\cite{Levin:2008mq}. The integer $z$ denotes the number of leaves in a complete periodic orbit, with larger values corresponding to more complex trajectories; $w$ represents the number of additional whirls the particle makes while moving outward toward the apoapsis; and $v$ characterizes the behavior of the next vertex the particle reaches after departing from the initial vertex (apoapsis)~\cite{Levin:2008mq}. Several studies on such zoom-whirl periodic orbits have been carried out in various gravitational theories~\cite{Levin:2008ci,Fujita:2009bp,Healy:2009zm,Wei:2019zdf,Azreg-Ainou:2020bfl,Deng:2020yfm,Deng:2020hxw,Wang:2022tfo,QiQi:2024dwc,Alloqulov:2025ucf,Haroon:2025rzx,Wang:2025hla,Lu:2025cxx}, including some quantum-corrected spacetimes within the framework of LQG~\cite{Tu:2023xab,Yang:2024lmj,Jiang:2024cpe,Al-Badawi:2025yum,Huang:2025vpi}.

Therefore, the main purpose of this paper is to investigate the periodic orbits of test particles and the GW radiation emitted from these orbits around the two quantum-corrected BH models proposed in~\cite{Zhang:2024khj}. Specifically, we will study the trajectory of particles by examining the geodesic equations for massive particles. Furthermore, we will analyze the effective potential of the massive particles to discuss their energy and angular momentum in bound orbits, and present the different types of periodic orbits around the two quantum-corrected BHs, studying the gravitational waveforms emitted from these periodic orbits. By analyzing the variations of periodic orbits and gravitational waveforms with respect to $\zeta$, we assess the possibility of distinguishing these two quantum-corrected BHs from the Schwarzschild BH.

The structure of this paper is organized as follows. In Section~\ref{section2}, we will briefly recall two static and spherically symmetric quantum-corrected BH models, and study the geodesics of massive particles around these BHs. In Section~\ref{section3}, we will further discuss the bound orbits of massive particles. By analyzing the effective potential for radial particle motion, we derive the allowed ranges of energy and angular momentum for these bound orbits. In Section~\ref{section4}, we will present different types of periodic orbits for massive particles, and discuss the influence of the quantum parameter $\zeta$ on these periodic orbits. In Section~\ref{section5}, we will investigate the gravitational waveforms emitted by periodic orbits, and examine the effect of $\zeta$ on the waveform characteristics. Finally, we provide a conclusion in Section~\ref{section6}. Throughout this paper, we will use geometric units with $ G = c = 1 $ and set the BH mass to be $ M = 1 $ unless stated otherwise.

\section{Timelike geodesics around quantum-corrected spacetime}\label{section2}

\begin{figure}[htbp]
	\centering
	\begin{subfigure}{0.45\textwidth}
		\includegraphics[width=3.2in, height=5.5in, keepaspectratio]{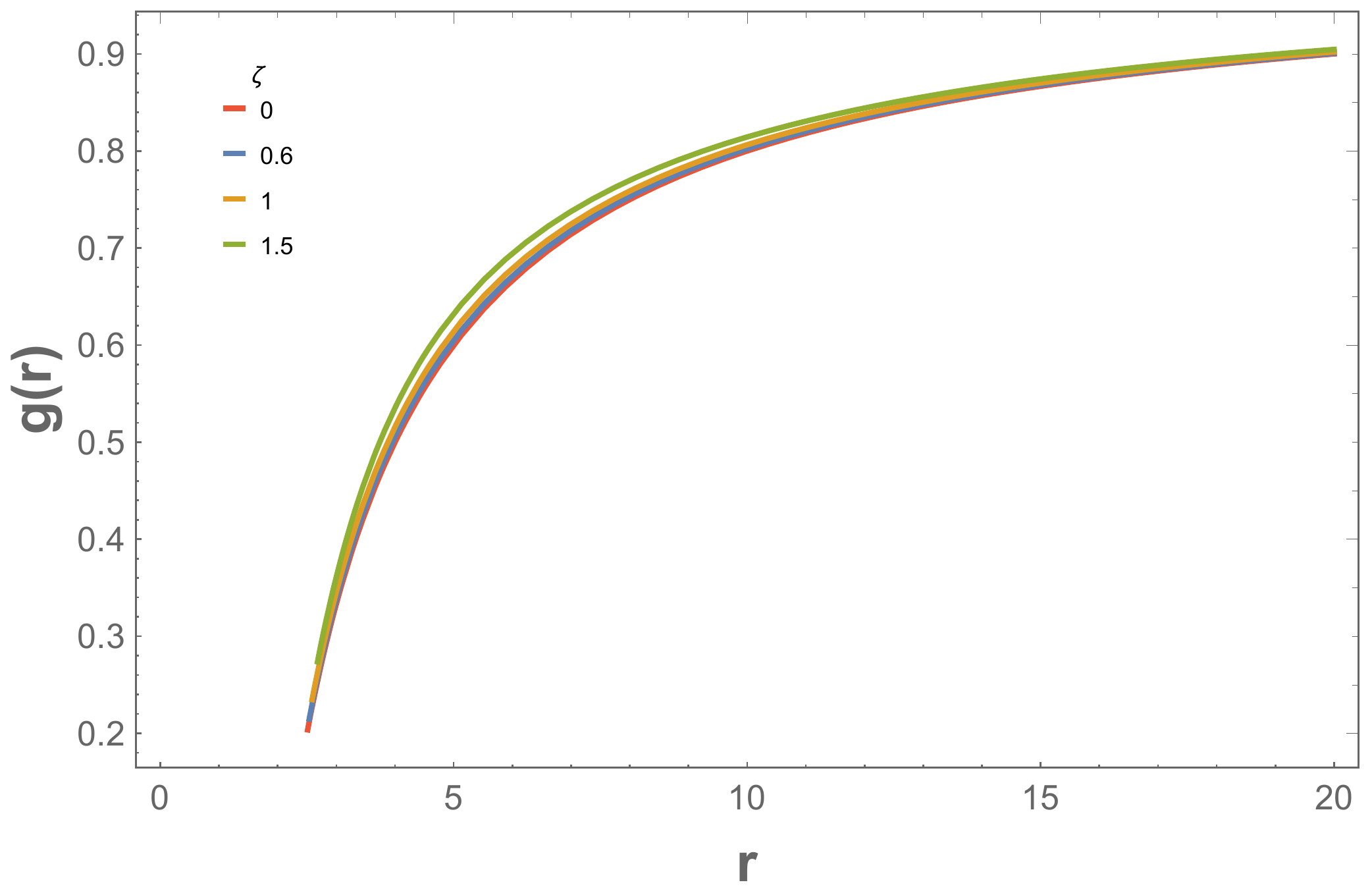}
	\end{subfigure}
	\caption{The metric function $g(r)$ for BH-I and BH-II with respect to $r$ with different values of $\zeta$.}
	\label{fig_function}
\end{figure}

In~\cite{Zhang:2024khj}, two static and spherically symmetric quantum-corrected BHs, denoted as BH-I and BH-II in the following, have been proposed in covariant approach to quantum gravity. The two quantum-corrected BH spacetimes can be described by the line element in Schwarzschild coordinates
\begin{equation}
	{\rm d} s^{2}=-f(r) {\rm d} t^{2}+ \frac{1}{g(r)} {\rm d} r^{2}+r ^2\left({\rm d} \theta^{2}+\sin ^{2} \theta {\rm d} \phi^{2}\right).\label{xianyuan}
\end{equation}
For BH-I, the metric functions read
\begin{equation}
 \begin{split}
 f(r)&=1-\frac{2 M}{r}+\frac{\zeta^{2}}{r^{2}}\left(1-\frac{2 M}{r}\right)^{2},\\
 g(r)&=1-\frac{2 M}{r}+\frac{\zeta^{2}}{r^{2}}\left(1-\frac{2 M}{r}\right)^{2},
 \end{split}
\end{equation}
and for BH-II, the metric functions are given by
\begin{equation}
	\begin{split}
		f(r)&=1-\frac{2M}{r},\\
		g(r)&=1-\frac{2 M}{r}+\frac{\zeta^{2}}{r^{2}}\left(1-\frac{2 M}{r}\right)^{2},
	\end{split}
\end{equation}
where $M$ and $\zeta$ denote the mass of BH and the quantum parameter, respectively. It is easy to see that both quantum-corrected BHs reduce to the Schwarzschild one as $\zeta \rightarrow 0$. Furthermore, in Fig.~\ref{fig_function}, we present the influence of $\zeta$ on the metric function $g(r)$ for BH-I and BH-II. It is found that the variation of $g(r)$ with $r$ for both qauntum-corrected BHs shows only slight differences compared to the Schwarzschild BH, indicating that the effect of $\zeta$ is weak.

Now we focus on the geodesic motion of a massive particle in the BH-I and BH-II spacetimes. For the massive particle confined to the equatorial plane ($\theta = \pi/2$), one has~\cite{Wald:1984rg,Liang:2023ahd}
\begin{equation}
	\begin{split}
		-1&=g_{ab}\left(\frac{\partial}{\partial \tau}\right)^a\left(\frac{\partial}{\partial \tau}\right)^b\\
		&=-f(r)\left(\frac{{\rm d} t}{{\rm d} \tau}\right)^2+\frac{1}{g(r)}\left(\frac{{\rm d} r}{{\rm d} \tau}\right)^2+r^2\left(\frac{{\rm d} \phi}{{\rm d} \tau}\right)^2,\label{keq1}
	\end{split}
\end{equation}
where $\tau$ is the affine parameter of the timelike geodesic of the particle, and $\left(\frac{\partial}{\partial \tau}\right)^a$ denotes the vector field tangent to the geodesic. It turns out that, in the static spherically symmetric spacetime described by Eq.~\eqref{xianyuan}, the energy $E$ and angular momentum $L$ per unit rest mass of the particle maintain conservation, which are determined by~\cite{Wald:1984rg,Liang:2023ahd}
\begin{equation}
 \begin{split}
 &E:=-g_{a b}\left(\frac{\partial}{\partial t}\right)^{a}\left(\frac{\partial}{\partial \tau}\right)^{b}=f(r) \frac{{\rm d} t}{{\rm d} \tau},\\
 &L:=g_{a b}\left(\frac{\partial}{\partial \phi}\right)^{a}\left(\frac{\partial}{\partial \tau}\right)^{b}=r^2 \frac{{\rm d} \phi}{{\rm d} \tau}.\label{EL1}
 \end{split}
\end{equation}
Inserting Eq.~\eqref{EL1} into Eq.~\eqref{keq1} yields
\begin{equation}
 \begin{split}
 \dot{r}^2\equiv\left(\frac{{\rm d} r}{{\rm d} \tau}\right)^2 =\left[E^2- V_\text{eff}(r;L)\right] \frac{g(r)}{f(r)},\label{ef}
 \end{split}
\end{equation}
where
\begin{equation}
	\begin{split}
		V_\text{eff}(r;L):=\left(1+\frac{L^2}{r^2}\right)f(r)\label{Veff}
	\end{split}
\end{equation}
is the effective potential associated to $L$. Note that in the case of $r \to \infty$, the effective potential $V_\text{eff}(r;L)$ in Eq.~\eqref{Veff} approaches $1$. Thus, in this case, it can be seen that from Eq.~\eqref{ef}, $E = 1$ serves as the upper bound of the energy for a bound orbit. Hence, the energy of a bound orbit should satisfy $E \leq 1$. The particle with $E > 1$ is on an unbound orbit and can escape to infinity.

By rewriting Eq.~\eqref{ef} with $\frac{{\rm d}r}{{\rm d}\tau} = \frac{{\rm d}r}{{\rm d}\phi} \frac{{\rm d}\phi}{{\rm d}\tau}$, the trajectory equation of the particle's motion can be obtained as
\begin{equation}
	\begin{split}
		\left(\frac{{\rm d} r}{{\rm d} \phi}\right)^2 =r^4 \left[\frac{{E}^2}{{L}^2}- \frac{V_\text{eff}(r;L)}{L^2}\right] \frac{g(r)}{f(r)}.\label{eq:motion}
	\end{split}
\end{equation}

\section{Bound orbits around quantum-corrected black holes}\label{section3}

In the previous section, the effective potential was introduced to determine the upper energy limit for bound orbits. In this section, the effective potential will be further used to define two particularly important types of bound orbit, namely the marginally bound orbit (MBO) and the innermost stable circular orbit (ISCO).

\begin{figure*}[htbp]
	\centering
	\begin{subfigure}{0.45\textwidth}
		\includegraphics[width=3in, height=5.5in, keepaspectratio]{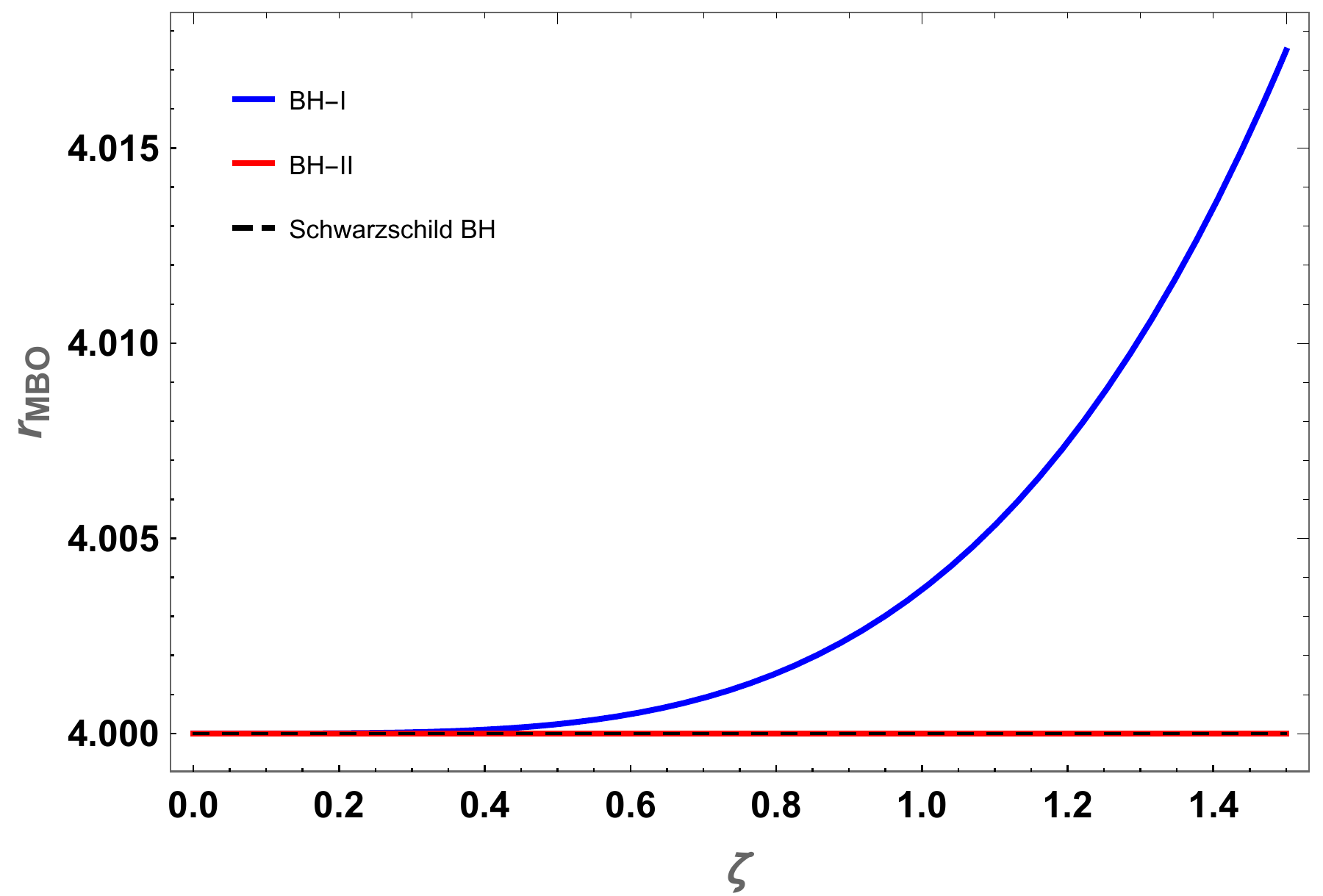}
	\end{subfigure}
	\hfill
	\begin{subfigure}{0.45\textwidth}
		\includegraphics[width=3in, height=5.5in, keepaspectratio]{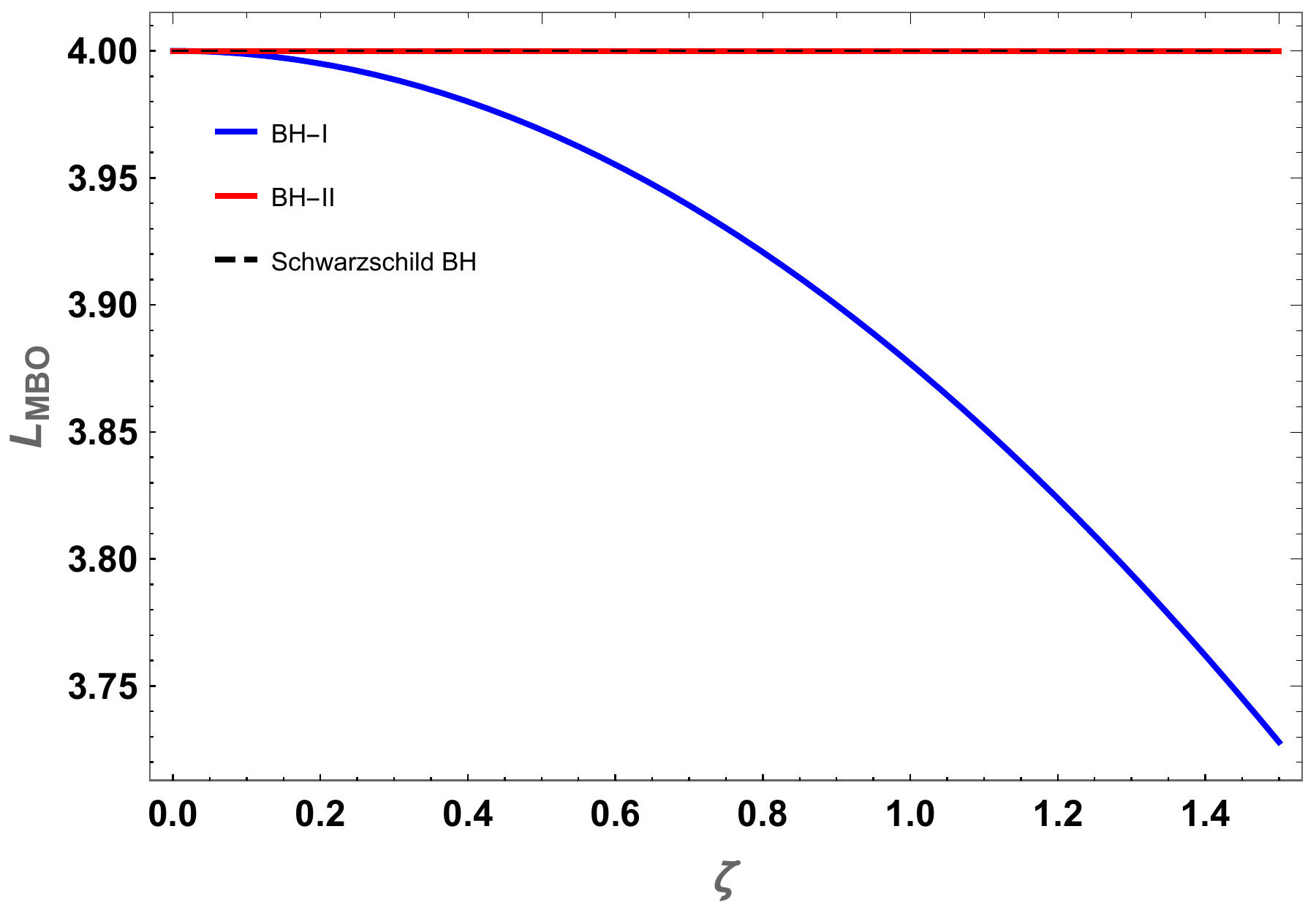}
	\end{subfigure}

		\caption{The variation in $r_{\rm MBO}$ and $L_{\rm MBO}$ with $\zeta$ for BH-I and BH-II. The blue solid line, red solid line, and black dashed line correspond to BH-I, BH-II, and Schwarzschild BH, respectively.}
	\label{fig:MBO}
\end{figure*}

As the innermost unstable bound orbit, the MBO, corresponding to the maximum energy ($E = 1$) attainable by a particle on a bound orbit, is defined by~\cite{Deng:2020yfm,Tu:2023xab}

\begin{equation}
	\begin{split}
		V_\text{eff}(r;L)= 1, \quad
		\frac{{\rm d} V_\text{eff}(r;L)}{{\rm d}r} = 0.\label{Veff1}
	\end{split}
\end{equation}
The solution to Eq.~\eqref{Veff1} will be denoted by $r_\text{MBO}$ and $L_\text{MBO}$. In Fig.~\ref{fig:MBO}, the variations of $r_{\rm MBO}$ and $L_{\rm MBO}$ with $\zeta$ for BH-I and BH-II are plotted. As illustrated in Fig.~\ref{fig:MBO}, for BH-I, increasing the values of the quantum parameter $\zeta$ leads to an increase in $r_{\rm MBO}$ and a decrease in $L_{\rm MBO}$. In contrast, for BH-II, the values of $r_{\rm MBO}$ and $L_{\rm MBO}$ remain independent of $\zeta$ and match the Schwarzschild case. In addition, the energy of the massive particle on the MBO always remains constant, i.e., $E_{\rm MBO} = 1$.

\begin{figure*}[htbp]
	\centering
	\begin{subfigure}{0.45\textwidth}
		\includegraphics[width=3in, height=5.5in, keepaspectratio]{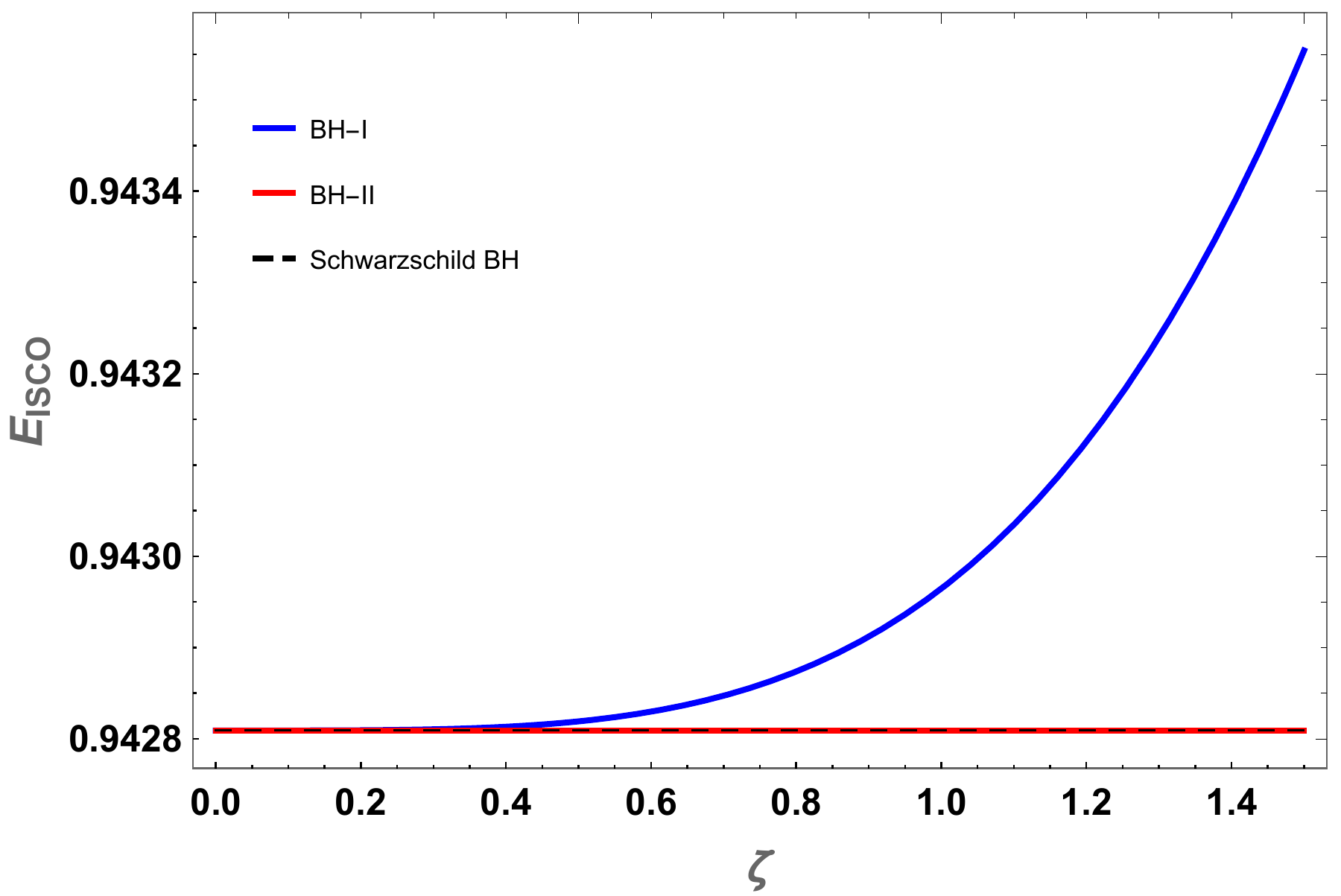}
	\end{subfigure}
	\hfill
	\begin{subfigure}{0.45\textwidth}
		\includegraphics[width=3in, height=5.5in, keepaspectratio]{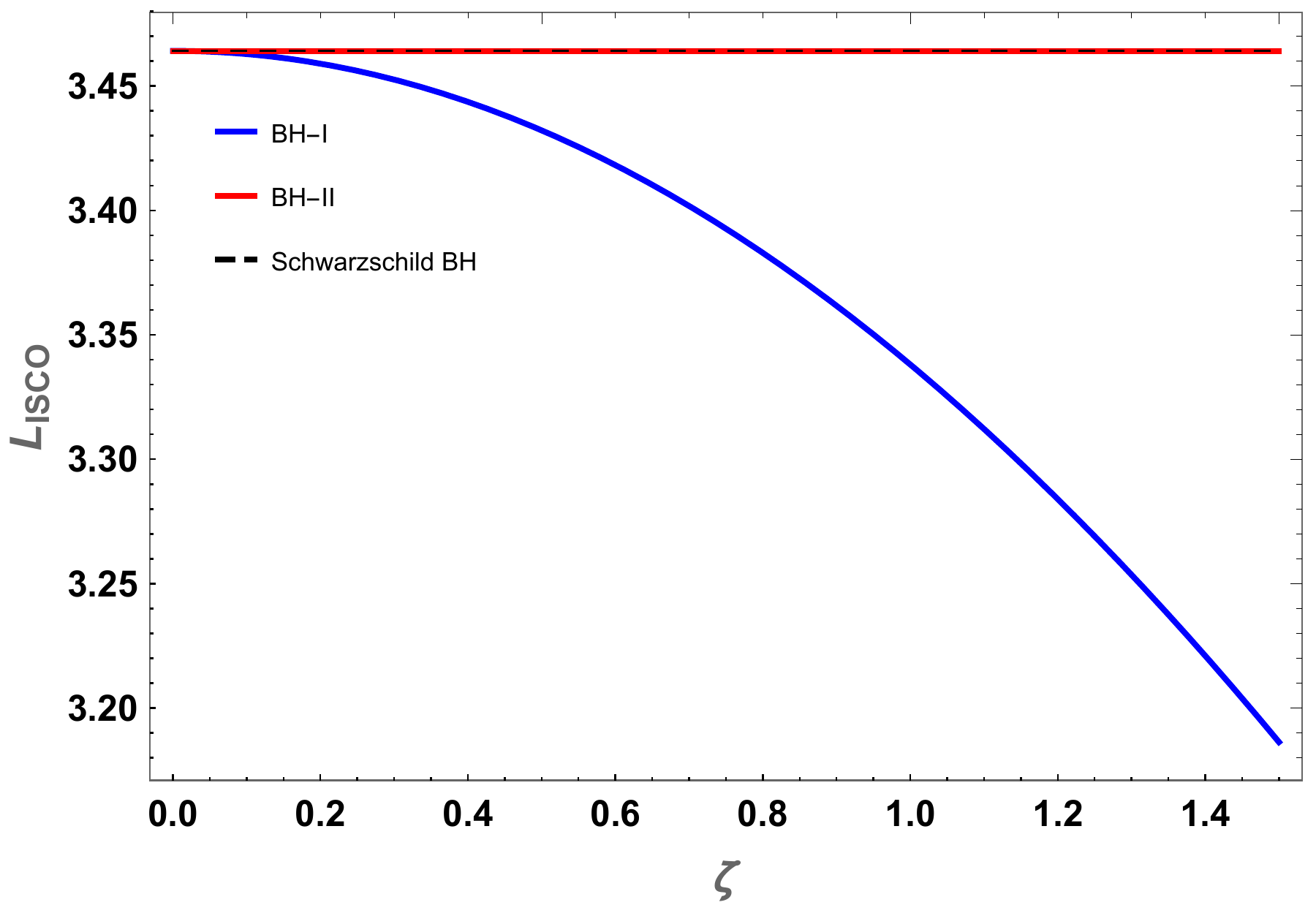}
	\end{subfigure}

	\caption{The behavior of $E_{\rm ISCO}$ and $L_{\rm ISCO}$ with $\zeta$ for BH-I and BH-II. The blue solid line, red solid line, and black dashed line correspond to BH-I, BH-II, and Schwarzschild BH, respectively.}
	\label{fig:ISCO}
\end{figure*}

On the other hand, the ISCO corresponds to the innermost stable circular orbit around a BH, which is determined by~\cite{Deng:2020yfm,Tu:2023xab}:
\begin{equation}
	\begin{split}
		V_\text{eff}(r;L)= E^2, \hspace{0.2cm}
		\frac{{\rm d} V_\text{eff}(r;L)}{{\rm d}r} = 0, \hspace{0.2cm}
		\frac{{\rm d}^2V_\text{eff}(r;L)}{{\rm d}r^2} = 0.
		\label{Veff2}
	\end{split}
\end{equation}
Solving Eq.~\eqref{Veff2}, we get the values of $r$, $L$ and $E$ on the ISCO as
\begin{equation}
	\begin{split}
		r_{\rm ISCO}&=\frac{3f(r_{\rm ISCO})f'(r_{\rm ISCO})}{2\left[f'(r_{\rm ISCO})\right]^2-f(r_{\rm ISCO})f''(r_{\rm ISCO})},\\
		L_{\rm ISCO}&=\sqrt{\frac{r_{\rm ISCO}^3 f'(r_{\rm ISCO})}{2 f(r_{\rm ISCO})-r_{\rm ISCO} f'(r_{\rm ISCO})}},\\
		E_{\rm ISCO}&=\frac{f\left(r_{\rm ISCO}\right)}{\sqrt{f\left(r_{\rm ISCO}\right)-\frac{1}{2} r_{\rm ISCO} f' \left(r_{\rm ISCO}\right)}},
		\label{eq:EL}
	\end{split}
\end{equation}
where the prime denotes the derivative with respect to $r$. Figure~\ref{fig:ISCO} illustrates the influence of $\zeta$ on $E_{\rm ISCO}$ and $L_{\rm ISCO}$, while the $\zeta$-dependence of $r_{\rm ISCO}$ has been established in prior work~\cite{Chen:2025ifv} and is not reproduced here. From Fig.~\ref{fig:ISCO}, it can be seen that for BH-I, both $E_{\rm ISCO}$ and $r_{\rm ISCO}$ increase with $\zeta$, while $L_{\rm ISCO}$ decreases monotonically as $\zeta$ increases. In contrast, BH-II remains unaffected under variations of $\zeta$.

\begin{figure}[htbp]
	\centering
	\begin{subfigure}{0.45\textwidth}
		\includegraphics[width=3.3in, height=5.5in, keepaspectratio]{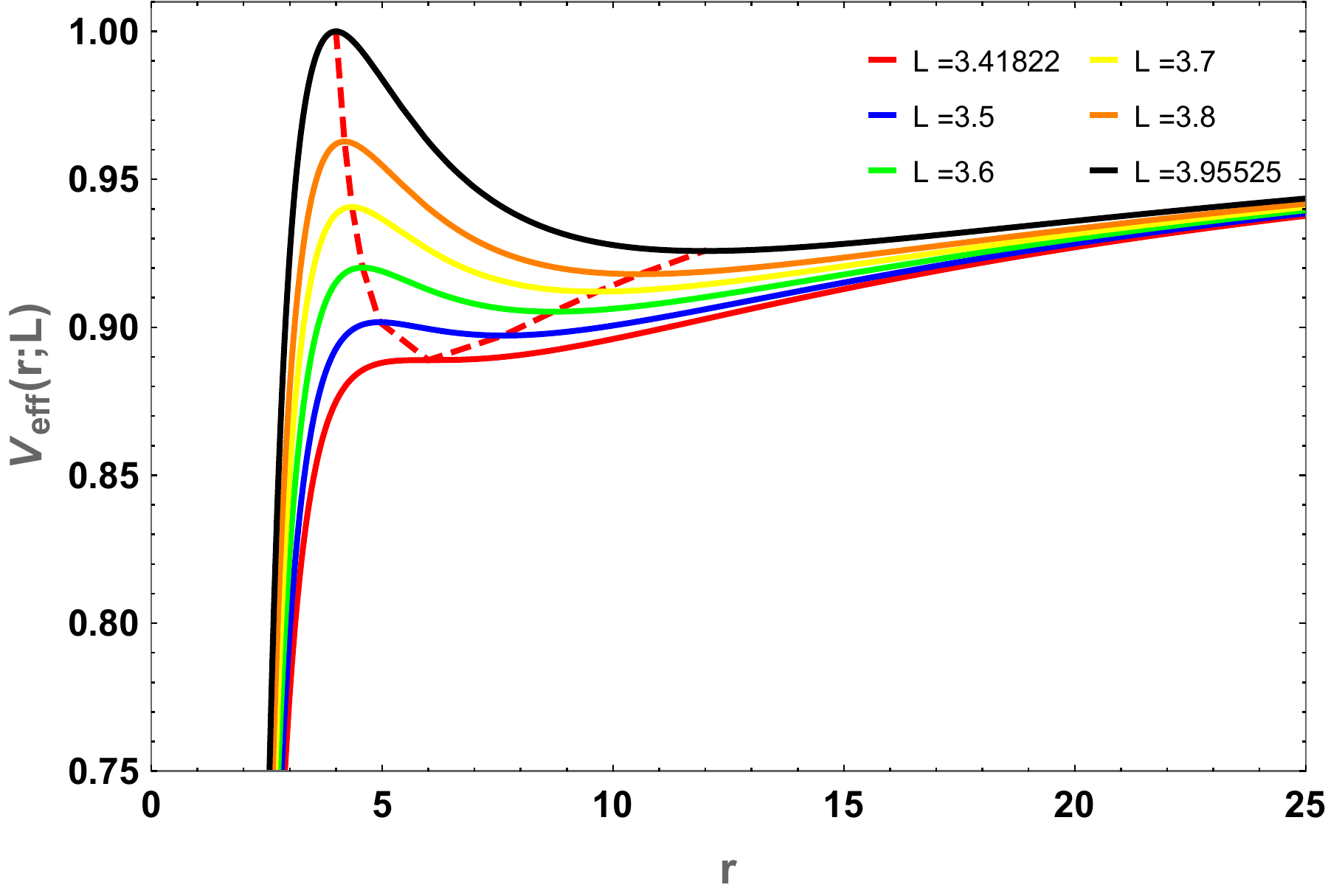}
	\end{subfigure}
	\caption{The effective potential $V_\text{eff}(r;L)$ as a function of $r$ with $L = 3.41822$, $3.5$, $3.6$, $3.7$, $3.8$, $3.95525$ for BH-I with $\zeta = 0.6$. The red dashed curve connects the extrema of these effective potentials. The red and black solid curves represent the effective potentials corresponding to the ISCO and MBO, respectively.}
	\label{fig_veff}
\end{figure}

As an example, Fig.~\ref{fig_veff} shows the effective potential as a function of $r$ with different $L$ for BH-I with $\zeta = 0.6$. In Fig.~\ref{fig_veff}, the red and black solid lines denote the effective potentials corresponding to $L_{\rm ISCO}=3.41822$ and $L_{\rm MBO}=3.95525$, respectively, while the red dashed line indicates the locations of the extrema of effective potential. It is evident that, except for the ISCO case where the effective potential features a single minimum, the effective potential exhibits two extrema, corresponding to stable and unstable circular orbits.

\begin{figure}[htbp]
	\centering
	\begin{subfigure}{0.45\textwidth}
		\includegraphics[width=3.2in, height=5.5in, keepaspectratio]{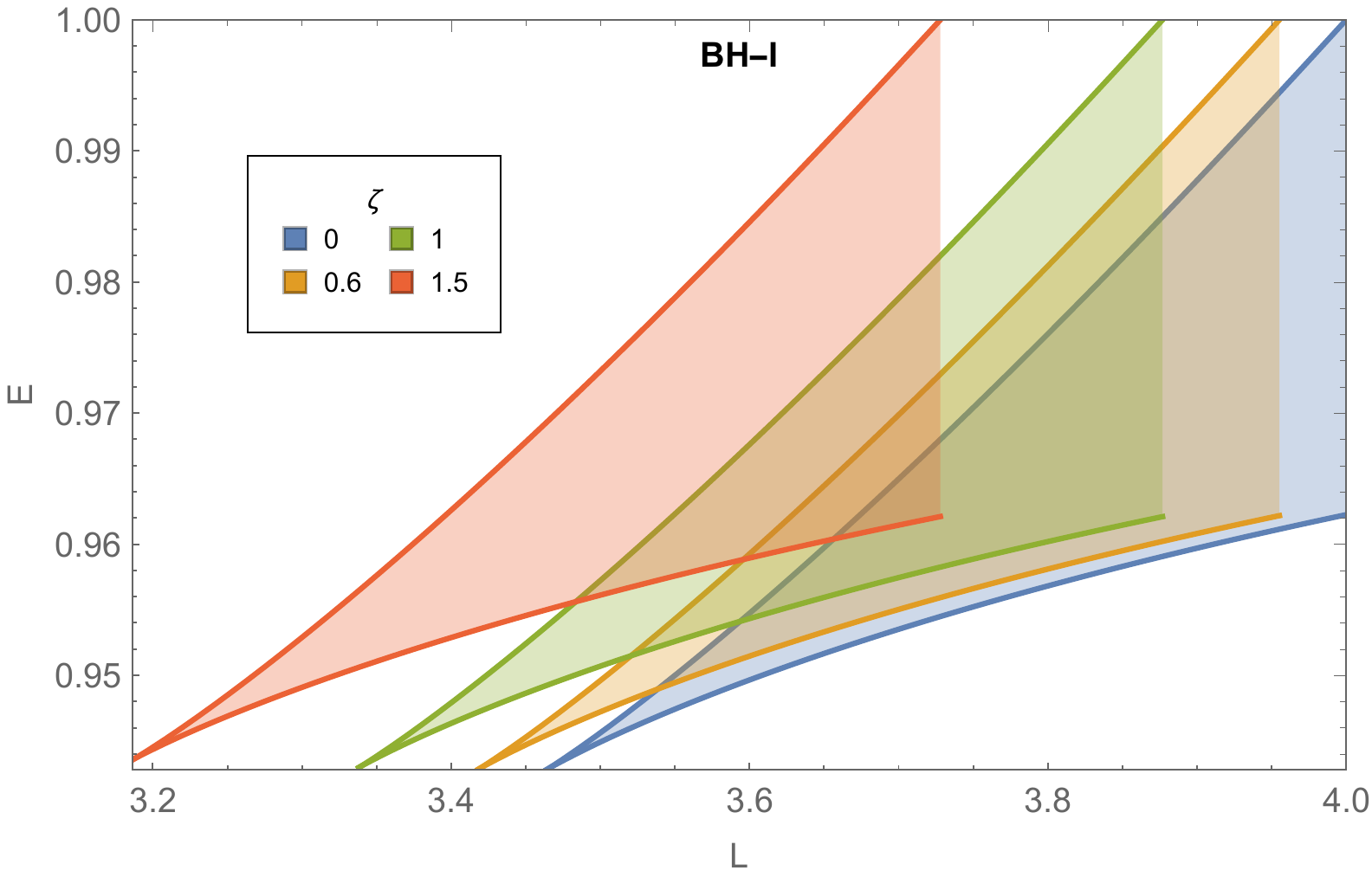}
	\end{subfigure}
	\caption{The allowed ranges of $E$ and $L$ for bound orbits around BH-I with different values of $\zeta$.}
	\label{fig_EL}
\end{figure}

It turns out that bound orbits are located between the MBO and ISCO with the energy $E$ satisfying $E_{\rm ISCO} \leq E \leq E_{\rm MBO}$~\cite{Deng:2020yfm,Tu:2023xab}. From Eq.~\eqref{ef}, it can be seen that $L$ and $E$ of the bound orbits should satisfy the requirement that $\dot{r}$ vanishes at some $r$. In Fig.~\ref{fig_EL}, we present the allowed ranges of energy and angular momentum for particles on bound orbits around BH-I with varying $\zeta$. The results demonstrate that as $\zeta$ increases, the bound orbit with a fixed $E$ can carry a smaller $L$, while the bound orbit with a fixed $L$ can take a greater $E$. Similarly, since the effective potential and the allowed $E$-$L$ range for BH-II always coincide with those of the Schwarzschild case, which are not presented in detail here.

\begin{figure}[htbp]
	\centering
	\begin{subfigure}{0.45\textwidth}
		\includegraphics[width=3.2in, height=5.5in, keepaspectratio]{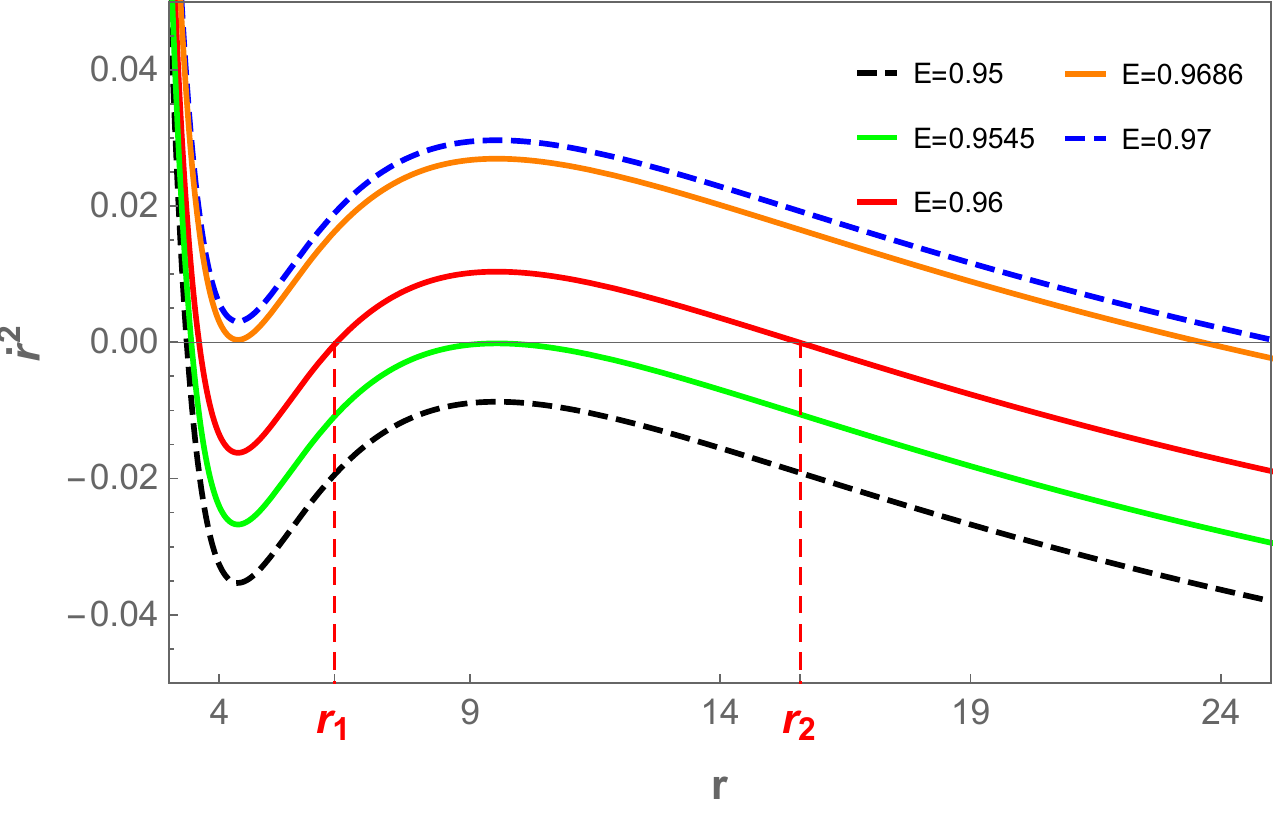}
	\end{subfigure}
	\caption{The root structure of $\dot{r}^2$ for BH-I with $\zeta = 0.6$ and $L = (L_{\rm ISCO} + L_{\rm MBO})/2$. The red solid line corresponds to bound orbit energy, green and orange lines indicate critical cases, and dashed lines represent unbound orbit energies.}
	\label{fig_veffyanshi}
\end{figure}

In Fig.~\ref{fig_veffyanshi}, we plot the square of the radial velocity, $\dot{r}^2$ in Eq.~\eqref{ef}, as a function of $r$ with a fixed $L$ in the BH-I spacetime. The red solid line corresponds to a bound orbit with an allowed $E$, on which $\dot{r}^2$ has at least two roots and the particle may move between the two roots ($r_1$ and $r_2$). The green and orange solid lines represent two critical cases, in which $\dot{r}^2$ has one root as its extremum. The dashed lines indicate the cases of non-bound orbits, in which $\dot{r}^2$ has at most one root.

\section{Periodic orbits of massive particles}\label{section4}

Among the bound orbits, periodic orbits, as a special class of bound orbits, hold significant importance in GW detection. The other bound orbits can be regarded as approximations of periodic orbits~\cite{Healy:2009zm}. For a particle moving along a periodic orbit, it oscillates between the two turning points $r_1$ (periastron) and $r_2$ (apastron) shown in Fig.~\ref{fig_veffyanshi} and returns to its initial position within a finite time. In this section, we explore periodic orbits around two quantum-corrected BHs.

Considering the spherical symmetry of the spacetime, we again focus on the motion of particles confined to the equatorial plane ($\theta = \pi/2$), where the dynamics are governed by the radial and azimuthal motions. It turns out that for a periodic orbit, the ratio of the radial oscillation frequency $\omega_r$ to the azimuthal frequency $\omega_\phi$ should be a rational number~\cite{Levin:2008mq}. Following~\cite{Levin:2008mq}, the rational number $q$ is related to $\omega_r$ and $\omega_\phi$ or three integers $(z, w, v)$ as
\begin{equation}
 q\equiv \frac{\omega_\phi}{\omega_r}-1 =\frac{\bigtriangleup\phi}{2\pi}-1= w + \frac{v}{z},
\end{equation}
where $z$, $w$, and $v$ represent the zoom, whirl, and vertex behaviors of the periodic orbits, respectively. The azimuthal angle $\bigtriangleup\phi$ over one complete radial cycle from apastron to the next apastron can be expressed as
\begin{equation}
	\bigtriangleup\phi = 2\oint {\rm d}\phi=2\int_{r_1}^{r_2}\frac{{\rm d}\phi}{{\rm d}r} {\rm d}r.\label{delta_fai}
\end{equation}
Combining Eqs.~\eqref{Veff}, \eqref{eq:motion}, and \eqref{delta_fai}, we obtain
\begin{equation}
	q=\frac{1}{\pi} \int_{r_1}^{r_2} \frac{1}{r^2 \sqrt{\left( \frac{E^2}{L^2}-\frac{f(r)}{L^2}-\frac{f(r)}{r^2}\right) \frac{g(r)}{f(r)}}}\,{\rm d}r-1.
\end{equation}
Hence, we have obtained the relationship between $q$ and the particle's energy and angular momentum.

\begin{figure*}[htbp]
	\centering
	\begin{subfigure}{0.45\textwidth}
		\includegraphics[width=3in, height=5.5in, keepaspectratio]{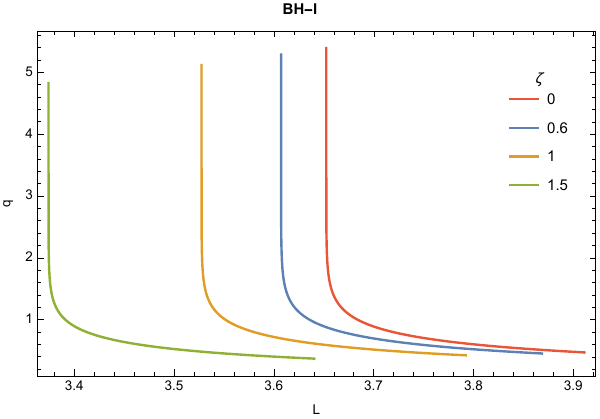}
	\end{subfigure}
	\hfill
	\begin{subfigure}{0.45\textwidth}
		\includegraphics[width=3in, height=5.5in,keepaspectratio]{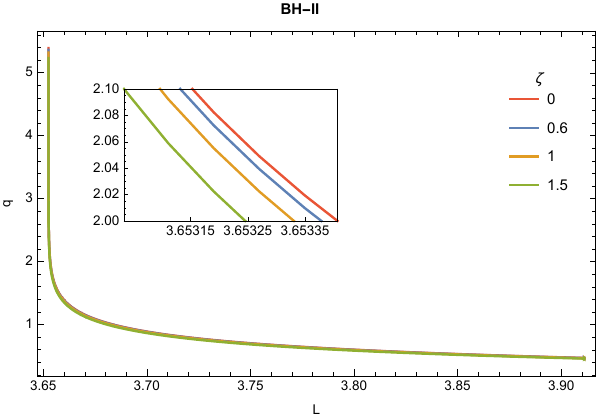}
	\end{subfigure}
	\begin{subfigure}{0.45\textwidth}
	\includegraphics[width=3in, height=5.5in, keepaspectratio]{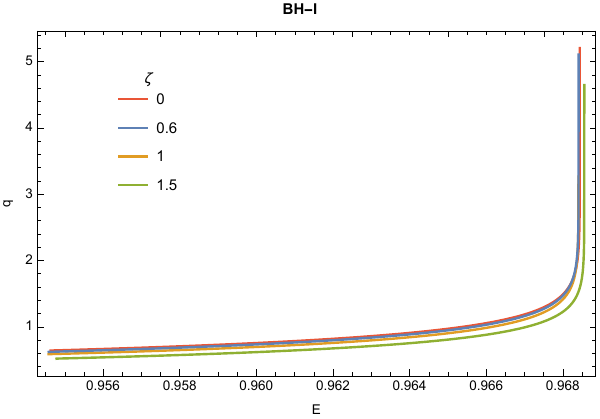}
	\end{subfigure}
	\hfill
	\begin{subfigure}{0.45\textwidth}
	\includegraphics[width=3in, height=5.5in, keepaspectratio]{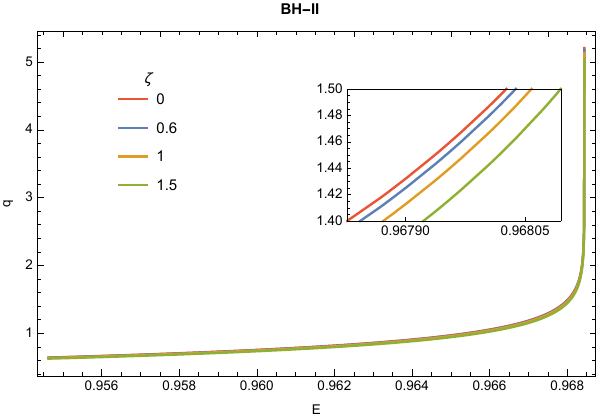}
	\end{subfigure}
	\caption{The variation of $q$ with $E$ and $L$ for BH-I and BH-II with different values of $\zeta$. The top panels display $q$ versus $L$ for BH-I (left) and BH-II (right) with $E=0.96$. The bottom panels show $q$ versus $E$ for BH-I (left) and BH-II (right) with $L=(L_{\rm ISCO} + L_{\rm MBO})/2$.}
	\label{fig:qEL}
\end{figure*}

In Fig.~\ref{fig:qEL}, we depict the variation of $ q $ with $ E $ and $ L $ for BH-I and BH-II with different values of $\zeta$. In the top two panels of Fig.~\ref{fig:qEL}, where the energy is fixed as $E = 0.96$, it is seen that the variation of $q$ with $L$ is similar for BH-I and BH-II. For BH-I, $q$ is significantly affected by $\zeta$ such that the entire $q$-$L$ curve shifts to the left as $\zeta$ increases. In contrast, BH-II is only slightly influenced by $\zeta$. This indicates that the presence of $\zeta$ allows the periodic orbits around BH-I to possess smaller angular momentum than those around a Schwarzschild BH. The bottom two panels of Fig.~\ref{fig:qEL}, where the angular momentum is fixed as $L = (L_{\rm ISCO}\; +\; L_{\rm MBO})/2$, show that the variation of $q$ with $E$ for both BH-I and BH-II is only weakly affected by $\zeta$.
Furthermore, Tables~\ref{tab:L} and \ref{tab:E} present the values of $E$ and $L$ for periodic orbits associated with different $(z,w,v)$. It can be seen that for BH-I, $\zeta$ has a more significant impact on $L$ than on $E$, whereas for BH-II, both $E$ and $L$ show minimal dependence on $\zeta$, aligning with the earlier discussion.

\begin{table*}[htbp]
	\centering
	\renewcommand{\arraystretch}{1.3}
	\setlength{\tabcolsep}{14pt}
	\begin{tabular}{|c|c|c|c|c|c|c|c|}
		\hline
		\multicolumn{2}{|c|}{\textbf{BHs}} &
		$L_{\scriptscriptstyle(1,1,0)}$ &
		$L_{\scriptscriptstyle(1,2,0)}$ &
		$L_{\scriptscriptstyle(2,1,1)}$ &
		$L_{\scriptscriptstyle(2,2,1)}$ &
		$L_{\scriptscriptstyle(3,1,2)}$ &
		$L_{\scriptscriptstyle(3,2,2)}$ \\
		\hline
		\multicolumn{2}{|c|}{\textbf{Schwarzschild}} &
		3.6835879 & 3.6534056 & 3.6575956 & 3.6527008 & 3.6553345 & 3.6526363 \\
		\hline
		\multirow{3}{*}{\textbf{BH-I}} & $\zeta=0.6$ & 3.6354747 & 3.6080248 & 3.6117462 & 3.6074172 & 3.6097260 & 3.6073657 \\ \cline{2-8}

		& $\zeta=1$ & 3.5512326 & 3.5281302 & 3.5311364 & 3.5276636 & 3.5294879 & 3.5276381 \\ \cline{2-8}

		& $\zeta=1.5$ & 3.3909453 & 3.3746143 & 3.3765842 & 3.3743511 & 3.3754841 & 3.3743714 \\
		\hline
		\multirow{3}{*}{\textbf{BH-II}} & $\zeta=0.6$ & 3.6830653 & 3.6533774 & 3.6574694 & 3.6526949 & 3.6552574 & 3.6526332 \\ \cline{2-8}

		& $\zeta=1$ & 3.6821561 & 3.6533296 & 3.6572530 & 3.6526851 & 3.6551257 & 3.6526285 \\ \cline{2-8}

		& $\zeta=1.5$ & 3.6804481 & 3.6532445 & 3.6568586 & 3.6526679 & 3.6548875 & 3.6526217 \\
		\hline
	\end{tabular}
		\caption{The values of $L_{(z,w,v)}$ for BH-I and BH-II with different values of $\zeta$, where $E=0.96$.}
	\label{tab:L}
\end{table*}

\begin{table*}[htbp]
	\centering
	\renewcommand{\arraystretch}{1.3}
	\setlength{\tabcolsep}{14pt}
	\begin{tabular}{|c|c|c|c|c|c|c|c|}
		\hline
		\multicolumn{2}{|c|}{\textbf{BHs}} &
		$E_{\scriptscriptstyle(1,1,0)}$ &
		$E_{\scriptscriptstyle(1,2,0)}$ &
		$E_{\scriptscriptstyle(2,1,1)}$ &
		$E_{\scriptscriptstyle(2,2,1)}$ &
		$E_{\scriptscriptstyle(3,1,2)}$ &
		$E_{\scriptscriptstyle(3,2,2)}$ \\
		\hline
		\multicolumn{2}{|c|}{\textbf{Schwarzschild}} &
		0.9654253 & 0.9683828 & 0.9680265 & 0.9684343 & 0.9682249 & 0.9684385 \\
		\hline
		\multirow{3}{*}{\textbf{BH-I}} & $\zeta=0.6$ & 0.9657035 & 0.9683629 & 0.9680504 & 0.9684067 & 0.9682254 & 0.9684099 \\ \cline{2-8}
		& $\zeta=1$ & 0.9661841 & 0.9683723 & 0.9681263 & 0.9684049 & 0.9682655 & 0.9684068 \\ \cline{2-8}
		& $\zeta=1.5$ & 0.9670765 & 0.9686368 & 0.9683886 & 0.9700572 & 0.9684674 & 0.9711449 \\
		\hline
		\multirow{3}{*}{\textbf{BH-II}} & $\zeta=0.6$ & 0.9654828 & 0.9683850 & 0.9680383 & 0.9684347 & 0.9682317 & 0.9684387 \\ \cline{2-8}
		& $\zeta=1$ & 0.9655823 & 0.9683888 & 0.9680584 & 0.9684354 & 0.9682434 & 0.9684390 \\ \cline{2-8}
		& $\zeta=1.5$ & 0.9657676 & 0.9683956 & 0.9680948 & 0.9684365 & 0.9682642 & 0.9684395 \\
		\hline
	\end{tabular}
	\caption{The values of $E_{(z,w,v)}$ for BH-I and BH-II with different values of $\zeta$ and a fixed $L =(L_{\rm ISCO} + L_{\rm MBO})/2$.}
	\label{tab:E}
\end{table*}

\begin{figure*}[htb]
	\centering
	\begin{subfigure}{0.33\textwidth}
		\includegraphics[height=5cm, keepaspectratio]{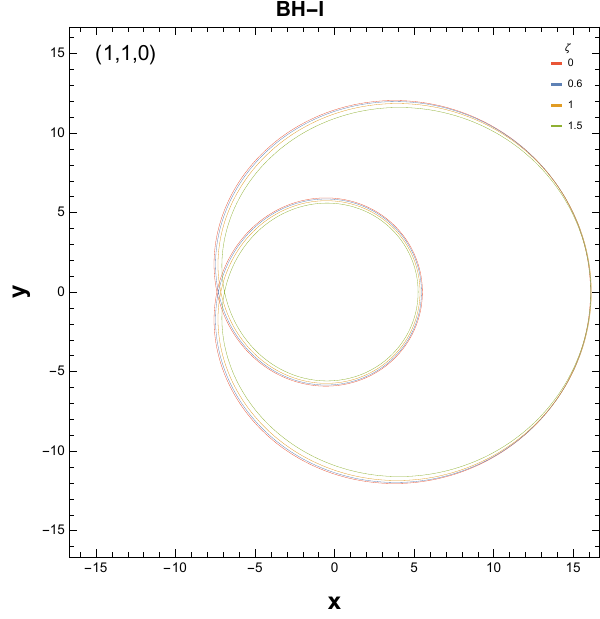}
	\end{subfigure}
	\begin{subfigure}{0.33\textwidth}
		\includegraphics[height=5cm]{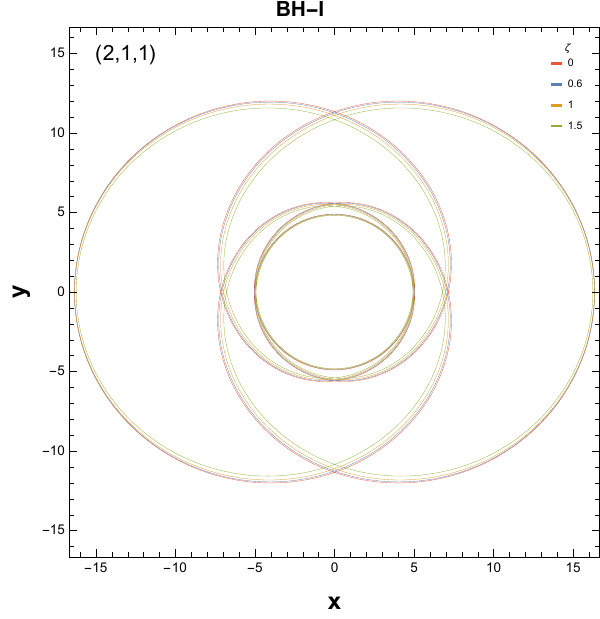}
	\end{subfigure}
	\begin{subfigure}{0.33\textwidth}
		\includegraphics[height=5cm,keepaspectratio]{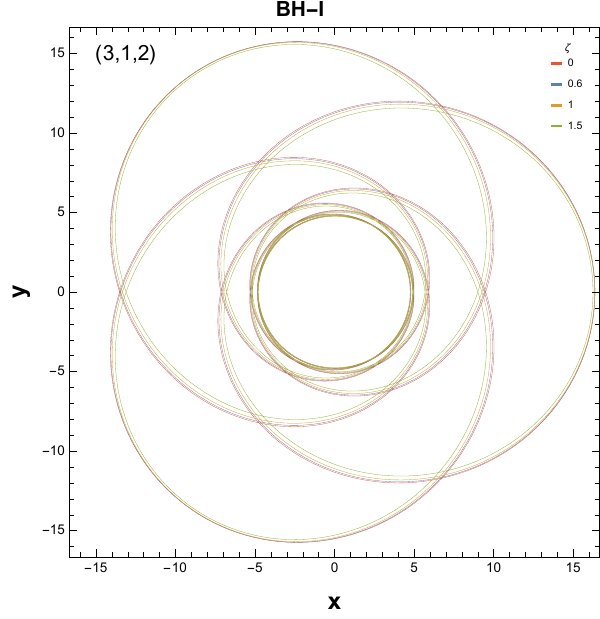}
	\end{subfigure}

	\begin{subfigure}{0.33\textwidth}
	\includegraphics[height=5cm, keepaspectratio]{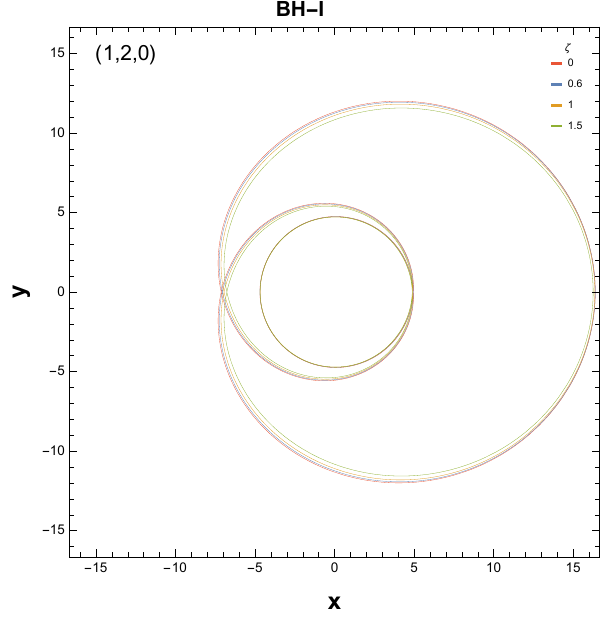}
	\end{subfigure}
	\begin{subfigure}{0.33\textwidth}
	\includegraphics[height=5cm]{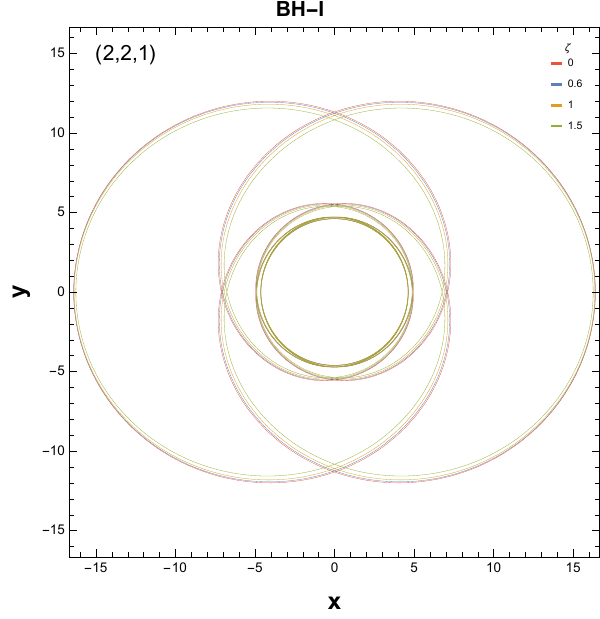}
	\end{subfigure}
	\begin{subfigure}{0.33\textwidth}
	\includegraphics[height=5cm,keepaspectratio]{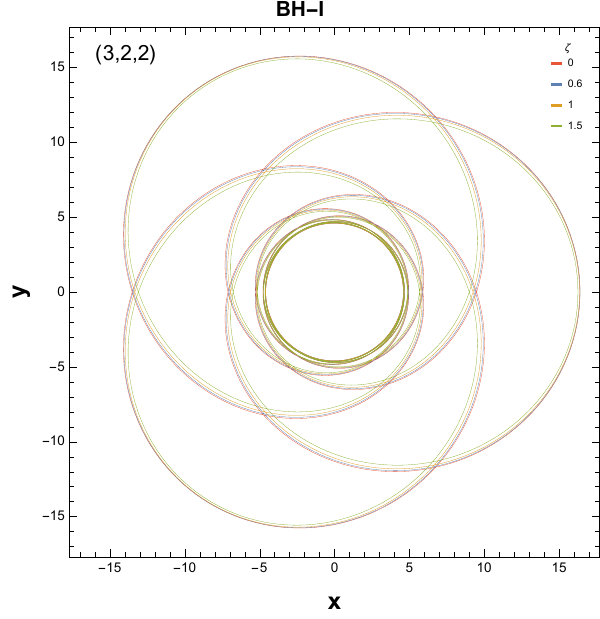}
	\end{subfigure}

	\caption{The periodic orbits characterized $(z,w,v)$ of BH-I with different values of $\zeta$, where $E=0.96$. The red, blue, orange, and green trajectories correspond to Schwarzschild BH, and BH-I with $\zeta=0.6$, $\zeta=1$, and $\zeta=1.5$, respectively.}
	\label{fig:zhouqi1}
\end{figure*}

\begin{figure*}[htb]
	\centering
	\begin{subfigure}{0.33\textwidth}
		\includegraphics[height=5cm, keepaspectratio]{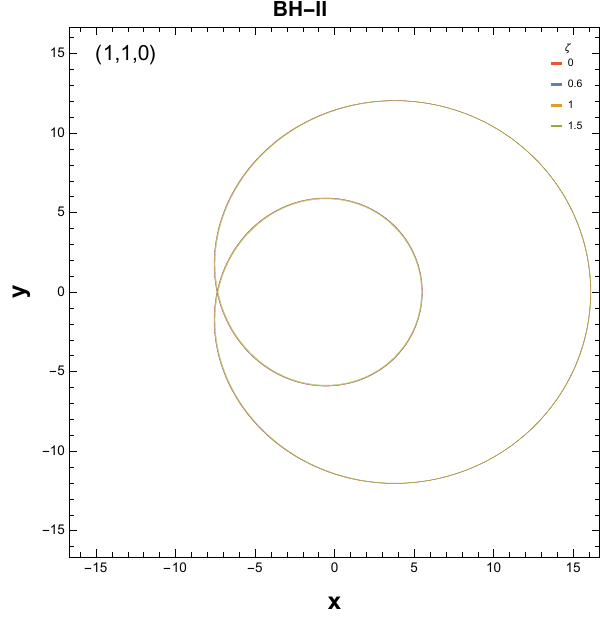}
	\end{subfigure}
	\begin{subfigure}{0.33\textwidth}
		\includegraphics[height=5cm]{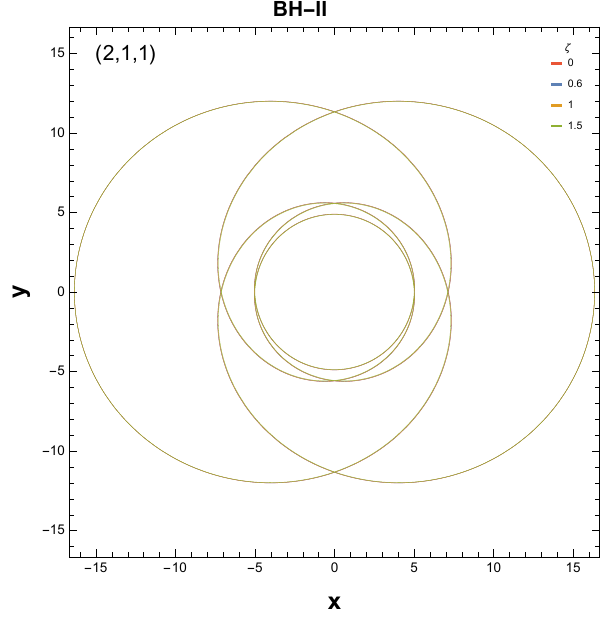}
	\end{subfigure}
	\begin{subfigure}{0.33\textwidth}
		\includegraphics[height=5cm,keepaspectratio]{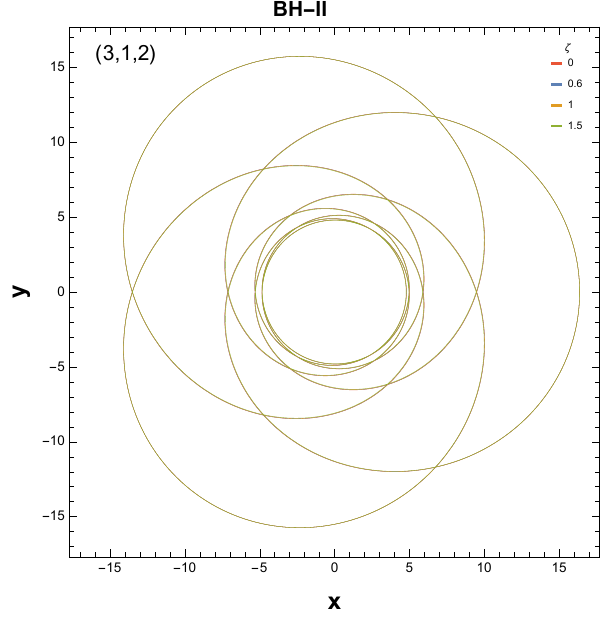}
	\end{subfigure}

	\begin{subfigure}{0.33\textwidth}
		\includegraphics[height=5cm, keepaspectratio]{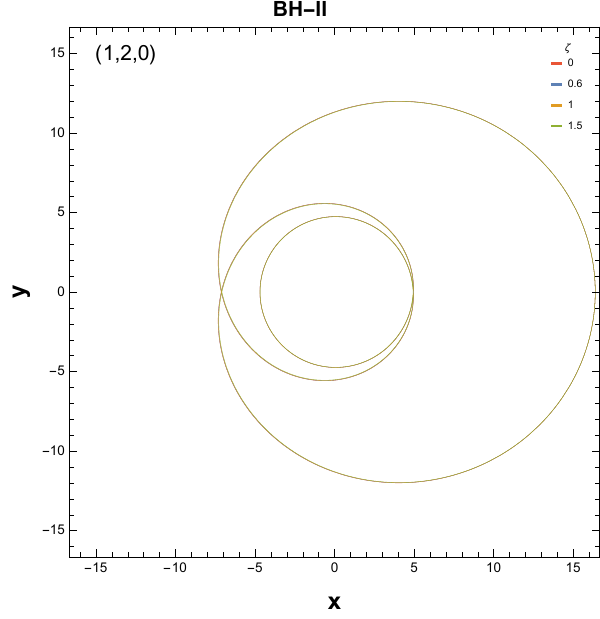}
	\end{subfigure}
	\begin{subfigure}{0.33\textwidth}
		\includegraphics[height=5cm]{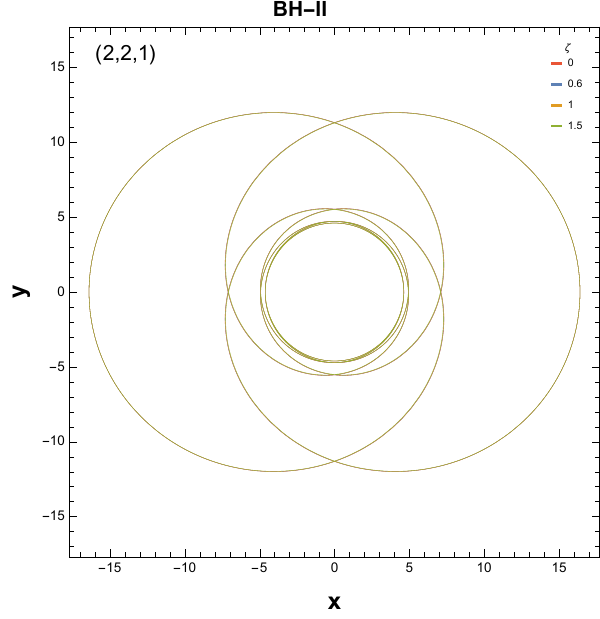}
	\end{subfigure}
	\begin{subfigure}{0.33\textwidth}
		\includegraphics[height=5cm,keepaspectratio]{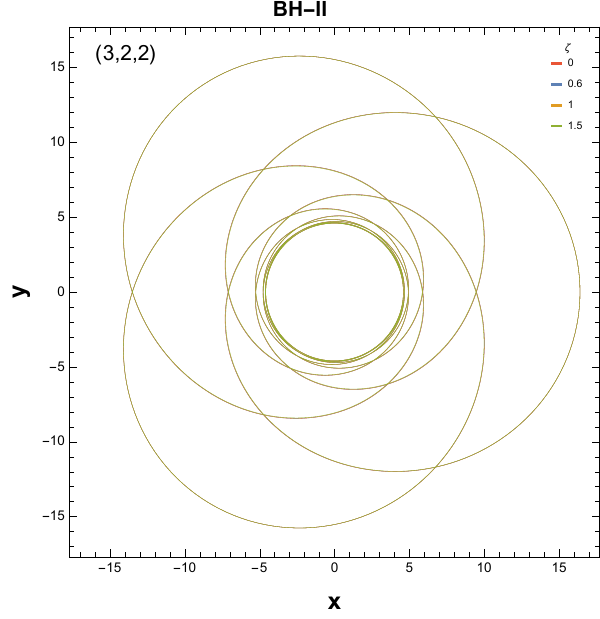}
	\end{subfigure}

	\caption{The periodic orbits $(z,w,v)$ of BH-II with different values of $\zeta$, where $E=0.96$. The red, blue, orange, and green trajectories correspond to Schwarzschild BH, and BH-II with $\zeta=0.6$, $\zeta=1$, and $\zeta=1.5$, respectively.}
	\label{fig:zhouqi2}
\end{figure*}

Based on the values of $E$ and $L$ listed in Tables~\ref{tab:L} and \ref{tab:E}, we numerically solve Eq.~\eqref{eq:motion} to obtain the relationship between the orbital radius $r$ and the azimuthal angle $\phi$, and present the resulting periodic orbits in the $x$-$y$ plane with $x:=r\cos\phi$ and $y:=r\sin\phi$. As an example, we fix $E=0.96$ and display the different periodic orbits characterized by $(z, w, v)$ around BH-I and BH-II in Figs.~\ref{fig:zhouqi1} and \ref{fig:zhouqi2}, respectively. From Figs.~\ref{fig:zhouqi1} and \ref{fig:zhouqi2}, comparing the periodic orbital results of BH-I and BH-II, we observe that both cases exhibit similarities with the Schwarzschild BH, exhibiting zoom-whirl characteristics. Moreover, we can see that the presence of $\zeta$ accelerates the variation of $r$ with $\phi$. Furthermore, the effect of $\zeta$ on the trajectories is more significant for BH-I, while for BH-II, the change is minor and the orbits remain almost indistinguishable from those of a Schwarzschild case.

\section{Gravitational waveforms from periodic orbits}\label{section5}

The GW radiation emitted by an EMRI system may encode rich information about the BH spacetime, and enables us to test modified theories of gravity in strong-field regimes. In what follows, we further investigate the GW radiation emitted by a test particle moving along periodic orbits around the two quantum-corrected BHs. It should be noted that GW radiation carries away energy and angular momentum from the EMRI system, leading to a time evolution of the orbital energy and angular momentum of the particle's periodic orbit~\cite{Bondi:1957dt}. However, since this evolution occurs sufficiently slowly over several orbital periods, the adiabatic approximation remains valid for studying GW radiation on such timescales~\cite{Drasco:2005is, Sundararajan:2008zm, Isoyama:2021jjd}.

Based on the periodic orbital results obtained in the previous section, we employ the kludge waveform model proposed in~\cite{Babak:2006uv} to investigate the GW radiation from test particles moving along periodic orbits around the two quantum-corrected BHs. For the EMRI system, the metric perturbation up to quadratic order is given by~\cite{Maselli:2021men,Liang:2022gdk}
\begin{equation}
 h_{ij} = \frac{4 G \eta M}{c^4 D_{\rm L}} \left(v_i v_j - \frac{G m}{r} n_i n_j\right).\label{hij}
\end{equation}
Here, $M$ and $m$ denote the BH mass and test particle mass respectively, $\eta = m M / (m+M)^2$ represents the symmetric mass ratio, and $D_{\rm L}$ is the luminosity distance of the EMRI system. The test particle's spatial velocity is given by $v_i$, while $n_i$ denotes the unit vector along its radial motion. It is worth noting that, since explicit numerical computations are required in this section, we no longer adopt the convention $G = c = M = 1$, but instead use their corresponding physical values. By projecting the GW signal onto the detector's coordinate system, the two GW polarization modes $h_+$ and $h_\times$, characterizing the gravitational waveforms of periodic orbits, are given by~\cite{Poisson:2014bk,Liang:2022gdk}
\begin{equation}
	\begin{split}
	h_{+} &= - \frac{2 \eta}{c^4 D_{\rm L}} \frac{(G M)^2}{r} \left(1+\cos^2\iota\right) \cos(2\phi+2 \omega), \\
	h_{\times} &= - \frac{4 \eta}{c^4 D_{\rm L}} \frac{(G M)^2}{r} \cos\iota \sin(2\phi+2 \omega),\label{jihua}
	\end{split}
\end{equation}
where $\iota$ denotes the inclination angle between the orbital angular momentum of the test particle and the observer's line of sight, while $\omega$ represents the longitude of the pericenter. The azimuthal angle $\phi$ and orbital radius $r$ of the particle's periodic motion are obtained by solving the geodesic equations~\eqref{EL1} and \eqref{ef}.

\begin{figure*}[htbp]
	\centering
	\begin{minipage}{0.48\textwidth}
		\centering
		\includegraphics[width=\textwidth]{BH1_zhouqi_312.pdf}
		\caption*{($a$)}
	\end{minipage}
	\hfill
	\begin{minipage}{0.48\textwidth}
		\centering
		\begin{minipage}{\textwidth}
			\includegraphics[width=\textwidth]{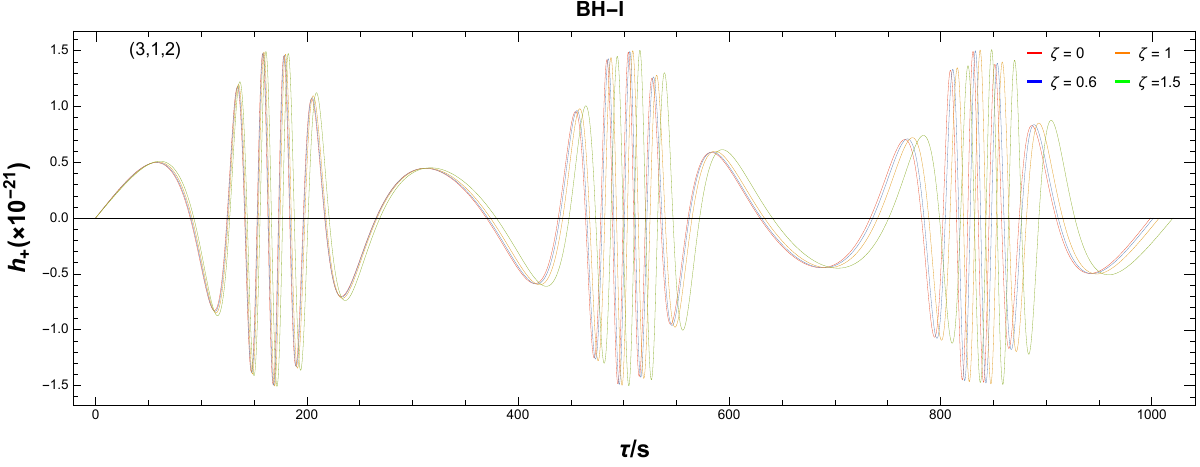}
			\caption*{($b$)}
		\end{minipage}
		\vspace{0.2cm}
		\begin{minipage}{\textwidth}
			\includegraphics[width=\textwidth]{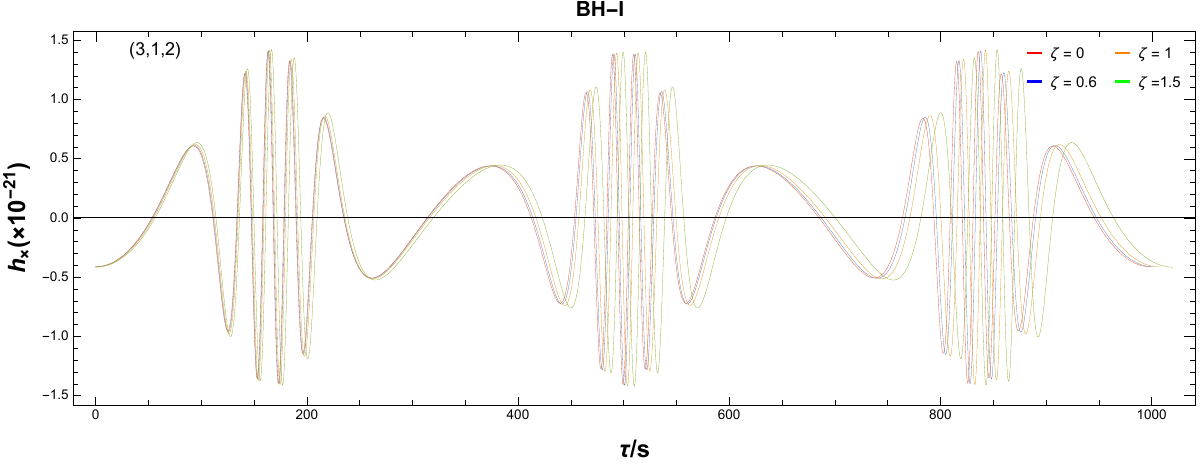}
			\caption*{($c$)}
		\end{minipage}
	\end{minipage}

	\caption{The gravitational waveform of a particle moving along the $(3,1,2)$ periodic orbit around BH-I with different values of $\zeta$. Panel ($a$) shows the periodic orbits, while panels ($b$) and ($c$) present the waveforms of $h_+$ and $h_\times$ as functions of proper time $\tau$($s$).}
	\label{fig_waveform}
\end{figure*}

\begin{figure*}[htbp]
	\centering
	\begin{minipage}{0.48\textwidth}
		\centering
		\includegraphics[width=\textwidth]{BH2_zhouqi_312.pdf}
		\caption*{($a$)}
	\end{minipage}
	\hfill
	\begin{minipage}{0.48\textwidth}
		\centering
		\begin{minipage}{\textwidth}
			\includegraphics[width=\textwidth]{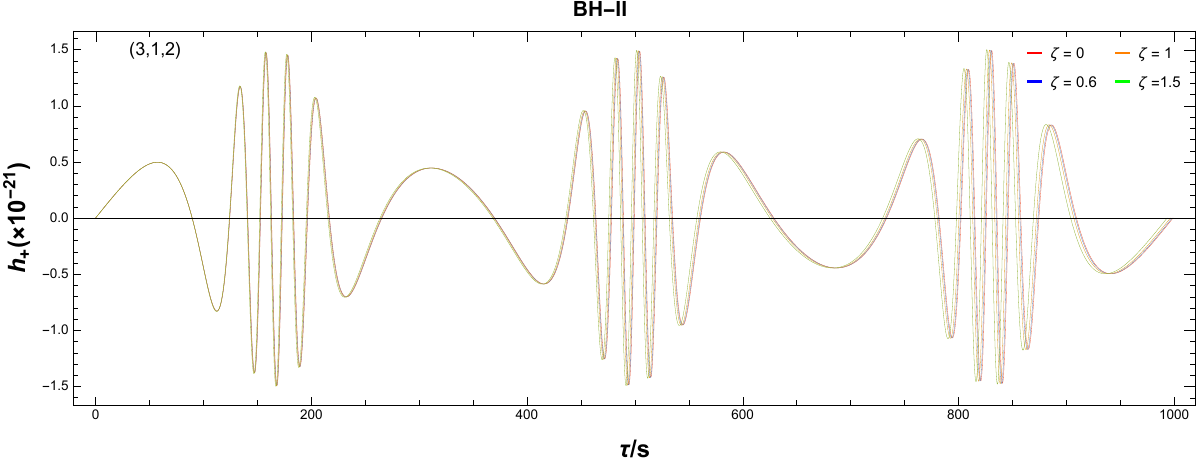}
			\caption*{($b$)}
		\end{minipage}
		\vspace{0.2cm}
		\begin{minipage}{\textwidth}
			\includegraphics[width=\textwidth]{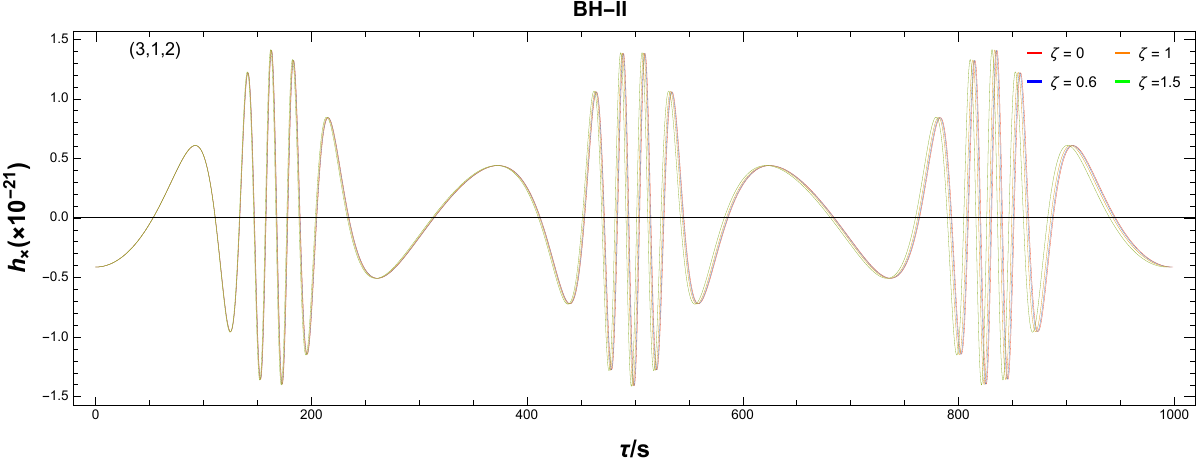}
			\caption*{($c$)}
		\end{minipage}
	\end{minipage}

	\caption{The gravitational waveform of a particle moving along the $(3,1,2)$ periodic orbit around BH-II with different values of $\zeta$. Panel ($a$) shows the periodic orbits, while panels ($b$) and ($c$) present the waveforms of $h_+$ and $h_\times$ as functions of proper time $\tau$($s$).}
	\label{fig_waveform1}
\end{figure*}

Given the parameters of the EMRI system composed of quantum-corrected BHs and a test particle, one can calculate the corresponding gravitational waveforms. We consider BHs with identical masses $M = 10^7 M_{\odot}$ and a test particle with mass $m=10 M_{\odot}$, where $M_{\odot}$ denotes the solar mass. The luminosity distance of the system is set to $D_{\rm L} = 200\; {\rm Mpc}$, while the inclination and longitude angles are set to $\iota = \pi/4$ and $\omega = \pi/4$, respectively. By solving the evolution of the phase angle with respect to the proper time $\tau$ and substituting it into Eq.~\eqref{jihua}, the gravitational waveforms for BH-I and BH-II can be obtained. In Figs.~\ref{fig_waveform} and \ref{fig_waveform1}, we present the waveforms corresponding to periodic orbits characterized by $(z,w,v) = (3,1,2)$ for different values of $\zeta$. It is evident that the gravitational waveforms for both BH-I and BH-II exhibit zoom and whirl behaviors corresponding to the periodic orbits within one orbital period. For BH-I, the parameter $\zeta$ significantly affects the GW phase such that the waveform shifts rightward $\zeta$ increases, causing a phase delay compared to the Schwarzschild waveform. This indicates prolonging of proper time $\tau$ per orbital period. In contrast, the waveform of BH-II shows minimal dependence on $\zeta$, leading to a subtle phase advance compared to the Schwarzschild BH as $\zeta$ increases. However, when $\zeta$ is small, the waveform for BH-II is nearly indistinguishable from that of the Schwarzschild one.

Moreover, we find that although the periodic orbits around BH-I and BH-II exhibit similar behavior with respect to variations in $\zeta$ as discussed in the previous section, the gravitational waveforms corresponding to these orbits show different behaviors as $\zeta$ varies. This is because the variation of the periodic orbits with respect to $\zeta$ is primarily determined by the behavior of $r(\phi)$, which exhibits similar $\zeta$-dependence for both BH-I and BH-II, leading to similar changes in the periodic orbits. In contrast, the gravitational waveforms depend on the variation of $r(\tau)$ and $\phi(\tau)$ with $\zeta$, which differ between BH-I and BH-II, thereby resulting in distinct waveform behaviors.

\section{Conclusion}\label{section6}

In this paper, we primarily study the orbital motion of massive particles around two quantum-corrected BHs. We first discussed the geodesic equations for particles moving in the equatorial plane. By analyzing the relative magnitudes of the particle's effective potential $V_{\rm eff}(r;L)$ and energy $E$, we determined the upper energy limit for the existence of bound orbits.

Then, we investigated two critical bound orbit types: the MBO and the ISCO. We discussed the dependence of their orbital radii ($r_{\rm MBO}$ and $r_{\rm ISCO}$) and angular momenta ($L_{\rm MBO}$ and $L_{\rm ISCO}$) on the quantum parameter $\zeta$. We found that, for BH-I, the orbital radii of both types of orbits increase with $\zeta$, while their angular momenta gradually decrease. In contrast, for BH-II, both the orbital radii and angular momenta are independent of $\zeta$, exhibiting behavior consistent with that of the Schwarzschild BH. Moreover, we analyzed the effective potential for bound orbits and determined the allowed range of energy and angular momentum for particles on these orbits. The results indicate that for BH-I, $\zeta$ significantly affects the allowed range, whereas for BH-II, the range remain unaffected by $\zeta$.

Furthermore, we studied the periodic orbits. The dependence of the rational number $q$ on the particle’s energy $E$, as well as its dependence on the angular momentum $L$, was shown in Fig.~\ref{fig:qEL}. Subsequently, in Tables~\ref{tab:L} and \ref{tab:E}, we provided the angular momenta $L_{(z,w,v)}$ of different periodic orbits characterized by $(z,w,v)$ with $E = 0.96$, and the energies $E_{(z,w,v)}$ with a fixed angular momentum $(L_{\rm MBO} + L_{\rm ISCO})/2$, respectively. Taking the periodic orbits with fixed $E=0.96$ as examples, we presented in Figs.~\ref{fig:zhouqi1} and \ref{fig:zhouqi2} the variations of different periodic orbits $(z,w,v)$ for BH-I and BH-II with respect to $\zeta$. It is observed that increasing the BH-I parameter $\zeta$ accelerates the variation of the particle’s orbital radius $r$ with azimuthal angle $\phi$, resulting in orbits with larger $\zeta$ lying closer to the BH than their Schwarzschild counterparts. The orbits around BH-II are also influenced by $\zeta$, but this effect is extremely weak and nearly indistinguishable from the Schwarzschild case.

Finally, we considered EMRI systems consisting of a test particle orbiting a supermassive BH (BH-I or BH-II), and investigated the GW radiation emitted from periodic orbit $(3,1,2)$ around both BH-I and BH-II. The corresponding waveform patterns were presented in Figs.~\ref{fig_waveform} and \ref{fig_waveform1}. It is found that for BH-I, the gravitational waveform exhibits an increasingly pronounced phase delay as $\zeta$ increases, causing the waveform to gradually deviate from that of the Schwarzschild case. For BH-II, increasing $\zeta$ leads to a slight phase advance in the waveform, but this effect remains minimal, making the waveforms almost indistinguishable from the Schwarzschild case when $\zeta$ is small.

In conclusion, our results demonstrate that the massive particle orbits around BH-I, as well as their corresponding gravitational waveforms, exhibit significant $\zeta$-dependent variations, leading to noticeable deviations from the Schwarzschild BH. While the periodic orbit around BH-II exhibits a weak dependence on $\zeta$, its gravitational waveform phase shows a slight advance as $\zeta$ increases. This phase advance in BH-II accumulates over time, causing the gravitational waveform to deviate slightly from the Schwarzschild case. When $\zeta$ is small for BH-II, this deviation in the waveform is almost indistinguishable from the Schwarzschild BH. However, the different phase variations with respect to $\zeta$ exhibited in the gravitational waveforms of BH-I and BH-II may also provide an important way to distinguish between these two quantum-corrected BHs in the future. Note that, in this work, we focus on the periodic orbits and GW radiation of massive particles around two quantum-corrected BHs in vacuum. Recently, the periodic orbits around BH-I in the presence of a quintessence field were studied in~\cite{Al-Badawi:2025yum}. An interesting extension of the study on periodic orbits and GW radiation in the vacuum case would be to include matter couplings, such as the coupling of electromagnetic field with cosmological constant, which will be left for future study.

\begin{acknowledgments}
This work is supported in part by NSFC Grant No. 12165005.
\end{acknowledgments}



\end{document}